\documentstyle[graphicx]{mn}

\title{Characterising the complex absorber in NGC 4151}

\author[N.\,J. Schurch and R.\,S. Warwick ]
{N.\,J. Schurch and R.\,S. Warwick \\
Department of Physics and Astronomy, University 
of Leicester, University Road, Leicester, LE1 7RH}
\date{}

\def\exo{{\it EXOSAT\/}}
\def\gin{{\it Ginga\/}}
\def\ro{{\it ROSAT\/}}
\def\asca{{\it ASCA\/}}
\def\ein{{\it EINSTEIN\/}}
\def\xte{{\it RXTE\/}}
\def\xmm{{\it XMM-Newton\/}}

\def\cha{{\it Chandra\/}}
\def\bep{{\it BeppoSAX\/}}
\def\int{{\it INTEGRAL\/}}
\def\cgro{{\it CGRO\/}}

\def\H0{{\rm ~km~s^{-1}~Mpc^{-1}}}

\def\etal{et al.~\/}

\def\la{\mathrel{\hbox{\rlap{\hbox{\lower4pt\hbox{$\sim$}}}{\raise2pt\hbox{$<$}}
}}}
\def\ga{\mathrel{\hbox{\rlap{\hbox{\lower4pt\hbox{$\sim$}}}{\raise2pt\hbox{$>$}}
}}}

\def\d25{D$_{25}$}

\def\.25{0.25 keV\thinspace}

\begin{document}

\maketitle

\begin{abstract}
We present a detailed analysis of the complex absorption apparent
in the 2--6 keV X-ray spectrum of the bright nearby Seyfert
galaxy NGC 4151. We first utilize the large bandpass and medium spectral 
resolution afforded by \bep ~data to construct a 1--100 keV spectral 
template, which assumes the absorption arises 
in both warm ({\it i.e.} partially  photoionized) and cold gas present
in the line of sight to the active nucleus of the source.
Application of this spectral model to an \asca ~``long-look'' observation
of NGC 4151 reveals a partial correlation between the underlying continuum 
flux and the ionization state of the warm absorber. Such a 
correlation is an intrinsic property of a warm absorber and argues strongly 
in favour of this interpretation for the complex absorbing column over 
alternative partial covering models. 

The warm absorber in NGC 4151 has a column density of $\sim 2 \times 10^{23} 
\rm~cm^{-2}$ with ionization parameter in the range log($\xi$) 
$\approx 2.4-2.7$. The inferred relatively low density 
($\sim 10^{5}$ cm$^{-3}$) for the warm gas,  implies an equilibration 
timescale 
for the dominant ions of the same order or longer than the timescale of the 
continuum variability. It follows that the warm component will
invariably be observed in a non-equilibrium ionization state. The warm 
absorber 
in NGC 4151 may be identified as a multi-temperature wind produced
by evaporation from the inner edge of an obscuring torus as discussed
in a recent paper by Krolik \& Kriss (2001). The unusually complex 
character of the absorption seen in NGC 4151 may then be explained
in terms of a fortuitous line of sight which grazes the top edge of the 
obscuring torus so as to intercept a substantial column of both the warm  
and cold gas.

We also find that (i) the reported hardening of the spectrum of NGC 4151 
as the continuum level falls may be simply due to the presence of an 
underlying (hard and relatively constant) Compton-reflection component and 
(ii) the iron K$\alpha$ line has a relatively narrow Gaussian profile and 
a line flux that remains constant over both short (days) and long 
(months to years) timescales - a relativistically broadened iron K$\alpha$ 
feature was not required in our modelling.

\end{abstract}


\section{Introduction}

The Seyfert 1 galaxy NGC 4151 was identified as an X-ray source 
over thirty years ago  (Gursky \etal 1971). As one of the brightest
Active Galactic Nuclei (AGN) accessible in the X-ray band, it has since been
extensively studied by missions such as \exo, \gin, \ro, \asca, \cgro
~and more recently \xte, \bep ~and  \cha. This observational
focus has revealed that the spectrum of NGC 4151 from 0.1--100 keV 
is comprised of a complex mixture of emission and absorption components, 
probably originating in a variety of locations from the innermost parts of 
the putative accretion disk in NGC 4151, out to the extended narrow-line 
region of the  galaxy. Given this spectral complexity it remains to be seen 
whether NGC 4151 should or should not be considered as an archetype of its 
class (Ulrich 2000; Zdziarski \etal 2001).

The intrinsic X-ray to $\gamma$-ray continuum emanating from the active 
nucleus of NGC 4151 appears to be produced by the thermal Comptonization 
of soft seed photons (e.g. Haardt \& Maraschi 1991; Zdziarski \etal 1994,
1996, 2000; Petrucci \etal 2000). There is also  
a contribution from reprocessing in the form of Compton-reflection 
and iron-K fluorescence components (e.g.  Maisack \etal 1991; Yaqoob \etal 
1995;  Zdziarski \etal 1996; 
Warwick \etal 1996; Piro \etal 2002; Zdziarski \etal 2001).  
Below $\sim$ 5 keV the 
hard continuum is strongly cut-off by photoelectric absorption in a 
substantial ($N_H \sim 10^{23} \rm~cm^{-2}$) line-of-sight gas column density  
(e.g. Holt \etal 1980; Yaqoob \etal 1993; Weaver \etal 1994a,b).
One of the long-standing problems in X-ray studies of NGC 4151 has been to 
understand the exact nature and origin of this absorption. X-ray
observations show very clearly that the cut-off is not as abrupt as one might
expected for absorption in a uniform slab of cold solar abundance material
(i.e., the increase in opacity with decreasing photon energy is less rapid
than predicted for a uniform cold absorber). Various solutions
have been suggested including an inhomogeneous cold absorber
(the partial covering model; Holt \etal 1980; Weaver \etal 1994b), 
an absorber with grossly non-solar abundances (Yaqoob \& Warwick 1991)
and the partial photoionization of the absorbing
medium (i.e., the warm absorber model, e.g., Krolik \& Kallman 1984). 
However, none of these
has provided a really satisfactory explanation of the nature 
of the absorber or the processes which give rise to the 
large changes in the absorption apparent in NGC 4151 on timescales 
of days or longer (Yaqoob \etal 1989, Yaqoob \etal 1993). 
Unfortunately the presence of additional soft X-ray emission components which
first appear at $\sim$ 2 keV and dominate the spectrum below $\sim$ 1 keV 
(Weaver \etal 1994a; Warwick, Smith \& Done 1995) adds further to the 
complication. \ein ~and \ro ~HRI measurements (Elvis, Briel \& Henry 1983;
Morse \etal 1995) and more recently observations by \cha ~(Ogle \etal 2000;
Yang, Wilson \& Ferruit 2001) 
have revealed that much of this soft emission emanates from a spatial 
resolved ($\sim 1.6$ kpc) highly ionized plasma coincident with the optical 
narrow-line region of the galaxy.

In the present paper we investigate the nature of the complex 
X-ray absorption column in NGC 4151 and in particular focus on the question 
of whether at least some of the properties of the absorber can be explained 
in terms of photoionization effects. Our approach has been to
use a high signal-to-noise \bep ~observation to define the parameters
of a ``spectral template'' representative of the underlying continuum 
and other features in the X-ray spectrum of NGC 4151. We then use this 
template to investigate the spectral variations which occurred in NGC 4151 
during a ``long-look observation'' carried out in the latter stages of the 
\asca ~mission. The plan of the paper is as follows. In \S 2 we specify 
our spectral template model and determine the relevant spectral parameters 
using the \bep ~measurements. In \S 3 we give brief details of the \asca ~
long-look observation, outline the data reduction techniques employed and 
characterize the flux and spectral variability apparent in the source
light curve. In the next section we describe the detailed spectral modelling 
of the \asca ~data focusing on the evidence for a photoionized absorber. 
Finally in \S 5 we discuss the implications of our results.

\section{The \bep ~data and the spectral template}

The broad bandpass of the instruments on \bep ~is a
tremendous help when attempting to constrain the form of the   
hard X-ray spectrum of NGC 4151. During the period 1996-1999, this source was 
observed three times with \bep ~as reported by Piro \etal (2002). Here we 
utilize the observation carried out in January, 1999, which provides the 
best signal to noise ratio of the available datasets.
X-ray data from three of the four instruments on \bep ~(namely the LECS, 
MECS and PDS instruments - see Parmar \etal 1997, Boella \etal 1997, 
Frontera \etal 1997 respectively were obtained from 
the ASI Science Data Center. In the case of the LECS and MECS 
instruments, source spectra were preprocessed via standard \bep ~procedures 
using on-source extraction regions of $6'$ and $4'$ radius respectively. 
Standard blank-sky background files were available for the LECS and MECS
whereas in the case of the PDS instrument, background-subtracted
source spectra were supplied directly. The spectra were accumulated over 
the full observation interval (see Piro \etal 2002) to give an on-source exposure time 
of 83 ks in the MECS and roughly half this for the other instruments.

A BL Lac object (MS 1207.9+3945) is located $\sim 5'$ to the north of 
NGC 4151. Piro \etal (2002) note that its count rate is $\sim 2\%$ 
and $\sim 20\%$ of that of NGC 4151 in the MECS and LECS instruments 
respectively. The contamination of the NGC 4151 spectra will be negligible 
for the MECS but may, on the basis of the Piro \etal (2002) figures, 
be as high as $\sim$20$\%$ in the LECS at $\sim 1$ keV. In 
practice the presence of this confusing source in the LECS data
will have little overall impact on our spectra fitting analysis 
(given the relatively soft spectrum of the BLLAC - Warwick, Smith 
\& Done 1995).

Spectral fitting was carried out using the {\bf XSPEC V11.0.1} software 
package with the Sept 1997 instrument response matrices.  For this purpose 
the source spectra were grouped to give a minimum of 20 counts per bin and 
the data below 1 keV (from the LECS) were excluded to avoid complications 
relating to the form of the soft X-ray emitting components and presence
of the soft confusing source.
We adopt a spectral template model which includes the following emission 
components:

\begin{enumerate}

\item  A power-law continuum with a normalization, $A_1$, and photon index, 
$\Gamma$, exhibiting a high-energy break at 100 keV;

\item A neutral Compton-reflection component (modelled by PEXRAV in XSPEC, 
Magdziarz \& Zdziarski 1995) with only the reflection scaling factor, $R$,
as a free parameter. The parameters relating to the incident continuum 
were tied to those of the hard power-law component. In
addition cos $i$ was fixed at 0.5 and the metal abundance in
the reflector was fixed at the solar value;

\item  An iron K$\alpha$ emission line of intensity $I_{K\alpha}$ at an energy
$E_{K\alpha}$ (with an intrinsic line width $\sigma_{K\alpha}$ set to 0.1 keV);

\item  A second power-law continuum representative of all the additional
soft X-ray emitting components with a slope coupled to that of the primary 
hard continuum but with a free normalization, $A_2$. 

\end{enumerate}

As noted earlier the complex absorption in NGC 4151 has in the past been 
modelled in number of ways, with arguably the most successful approach being 
the partial covering (sometimes referred to in the literature as a dual 
absorber) model (e.g. Holt \etal 1980; Yaqoob \& Warwick 1991; 
Weaver \etal 1994b). In the partial covering scenario a fraction $f_{cov}$ 
of the hard continuum is absorbed by a cold gas column density $N_{H,1}$,
whilst the remaining  $1-f_{cov}$ fraction of the continuum intercepts a 
reduced column density $N_{H,2}$.  In the present paper we prefer to focus 
on a model 
employing the same number of free parameters but with the absorber stratified 
along the line of sight rather than perpendicular to it. In our preferred 
scenario the complex absorber is represented as product of two absorption 
components, namely a warm column density $N_{H,warm}$ and a cold gas column 
$N_{H,cold}$. For the former we use multiplicative table models in XSPEC 
generated via the photoionization code XSTAR (Kallman 2001). The ionization 
state of the warm gas is governed by the ionization parameter 
$\xi=L_{ion}/n~r^{2}$ where $L_{ion}$ is the source luminosity
in the 0.0136-13.6 keV bandpass in $\rm~erg~s^{-1}$, $n$ is the number of
hydrogen atoms/ions in the gas per $\rm~cm^{-3}$ and $r$ is the distance from 
the central source to the inner edge of the warm cloud in cm.  We assumed an 
ionizing continuum of the form adopted by Krolik \& Kriss (1995). For further
details of the photoionization modelling see Griffiths \etal (1998). The 
cold absorption was represented by the {\bf wabs} model in XSPEC which is 
based on 
the absorption cross-sections tabulated in Morrison \& McCammon (1983).  
Here we assume solar abundances in both the warm and cold absorber. 

The adopted spectral model assumes that {\it only} the hard power-law 
continuum is subject to the complex absorption. However absorption 
arising in the line-of-sight column density through our own Galaxy is 
applied to all four emission components ($N_{H,Gal} = 2 \times 10^{20} 
\rm~cm^{-2}$).

The results of fitting this spectral template to the \bep ~data are detailed in 
Table~\ref{bsax4-1}, which lists the best fitting values for the nine 
free parameters of the model. The quoted errors (here as elsewhere in this 
paper) are at the  90\% confidence level as defined by a 
$\Delta$$\chi$$^{2}$=2.71 criterion ({\it i.e.} assuming one interesting 
parameter).

\begin{table}
\centering
\begin{minipage}{85 mm} 
\centering
\caption{The results of fitting the ``spectral template'' to the 
\bep ~observation}
\begin{tabular}{lll}

Model & Best-Fit & Units \\
Parameter & Value & \\ \hline 

N$_{H,warm}$ & 23.9$^{+0.9}_{-0.9}$ & $10^{22}\rm~cm^{-2}$ \\
N$_{H,cold}$ & 3.4$^{+1.1}_{-0.8}$  & $10^{22}\rm~cm^{-2}$ \\ 
$\Gamma$ & 1.65$^{+0.02}_{-0.03}$ &\\
R & 0.37$^{+0.18}_{-0.15}$ & \\
log($\xi$) & 2.48$^{+0.06}_{-0.03}$ &\\ 
$A_1$ & 4.6$^{+0.2}_{-0.2}$ & $10^{-2}\rm photon~keV^{-1}~cm^{-2}~s^{-1}$  \\ 
$A_2$ & 1.8$^{+0.1}_{-0.1}$& $10^{-3}\rm photon~keV^{-1}~cm^{-2}~s^{-1}$\\
$E_{K\alpha}$ & $6.37^{+0.03}_{-0.02}$ & keV\\ 
$I_{K\alpha}$ & $3.4^{+0.4}_{-0.3}$ & $ 10^{-4} \rm photon~cm^{-2}~s^{-1}$ \\
 & \\
$\chi^{2}$ & 832 \\
d-o-f & 815\\ \hline
\end{tabular}
\label{bsax4-1}
\end{minipage}
\end{table}

\begin{figure}
\centering
\begin{minipage}{85 mm} 
\centering \hbox{ \includegraphics[width=5.5 cm, 
angle=270]{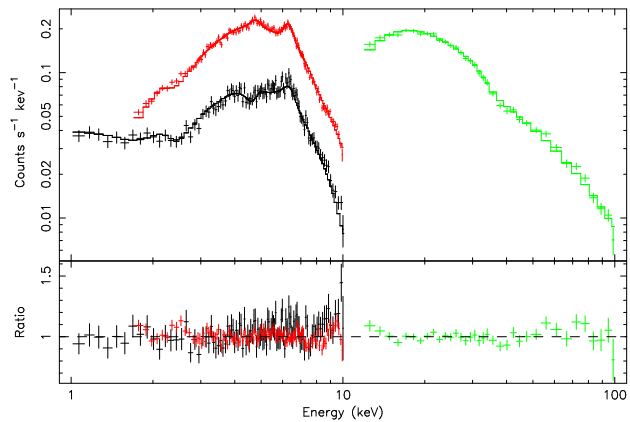}}
\caption{The \bep ~LECS (black), MECS (red) and PDS (green) data fitted by our spectral template model. 
{\it Upper panel:} The count rate spectra and best-fit model. {\it Lower panel:} 
The ratio of the data to the best-fit model.}
\label{Beppo}
\end{minipage}
\end{figure}

\begin{figure}
\centering
\begin{minipage}{85 mm} 
\centering \hbox{ \includegraphics[width=6 cm, 
angle=270]{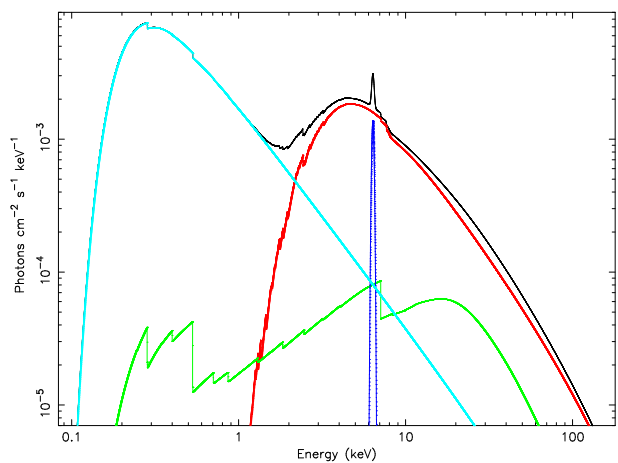}}
\caption{The spectral template model giving the best-fit to the \bep ~data.
There are four emission components comprising of a hard power-law
continuum (red), Compton reflection (green), an iron K$\alpha$ line (dark blue) and a second
power-law representing soft X-ray emitting components (light blue). Complex 
absorption (warm+cold components) causes the low-energy cut-off in the hard 
continuum. All the emission components are subject to Galactic
absorption.}
\label{model}
\end{minipage}
\end{figure}

The spectral template provides an excellent fit to the \bep ~data
as illustrated in Fig. \ref{Beppo}. Fig. \ref{model} shows the 
corresponding best-fitting model spectrum.

\section{The \asca ~long-look observation}

In the present paper we concentrate on the final observation of NGC 4151 
carried out by \asca ~during the period May 12-25, 2000. This was an 
exceptionally long observation carried out during the last phase of the
\asca ~programme, fully meriting its description as a ``long-look''. 
The \asca ~payload included four X-ray telescopes, two equipped with 
Solid-state Imaging Spectrometers (SIS) and two with Gas 
Imaging Spectrometers (GIS).
All the relevant datasets pertaining to the SIS and GIS instruments were 
obtained from the \asca ~public archive at the HEASARC. Standard
screening criteria and reduction techniques were employed using software
routines within the FTOOLS package. During the long-look observation 
the SIS instruments were operated in a single CCD mode with the nearby
BL Lac object positioned at the very edge or just off-chip in the SIS-0
and SIS-1 instruments respectively. In an attempt to reduce both the
impact of the instrument background and the contamination from the BL Lac
to negligible proportions, we applied a $1'$ radius source extraction
circle to the SIS data. Compared to a more standard $3'$ radius extraction 
cell, this reduced the 1-10 keV SIS count rate by $\sim 11\%$. For 
the GIS data we used a $3'$ radius source extraction region and a background 
spectrum taken from an off-source region in the GIS field of view.

As discussed recently by Turner \etal (2001) the \asca ~SIS detectors
have shown a degradation in efficiency at lower energies, which is probably
due to an increased dark current levels and decreased charge transfer 
efficiency (CTE), producing SIS spectra which diverge from each other and
from the GIS data. Furthermore, data from the last phase of \asca ~operations 
(AO-8) have revealed a non-linear evolution of the SIS CTE. Here we have
applied the interim solution released on 2001 February 13 by the \asca ~GOF 
(in the form of the CTE file $ sisph2pi_{-}130201.fits $) which reduces some 
but not all of the spectral inconsistencies (see below). 
We restrict our analysis to the 1--10 keV band data to avoid the 
worst of the low energy calibration uncertainties in the SIS instruments 
(e.g. see Appendix A of Weaver \etal 2001).  For subsequent analysis,
the total accumulated on-source exposure was 285 ks in the SIS and 435 ks
in the GIS detectors.

\begin{figure}
\centering
\begin{minipage}{85 mm} 
\centering\hbox{\includegraphics[width=6 cm, angle=270]{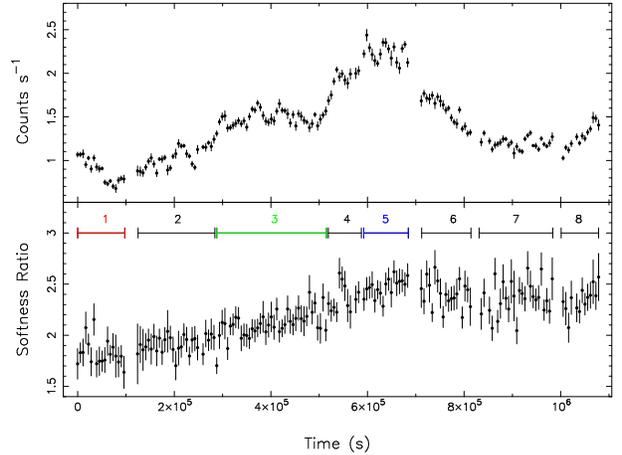}}
\caption{{\it Upper panel:} the combined SIS 1-10 keV X-ray light curve from 
the ``long-look' \asca ~observation. {\it Lower panel:} 
The variation the 2--5 keV/5--10 keV softness ratio during the observation.
The horizontal bars illustrate how the observation was split into
the eight time segments for the spectral analysis. The coloured horizontal bars indicate the segments for which spectra are shown in Fig. \ref{spectra}.
}
\label{light}
\end{minipage}
\end{figure}

The light curve in the full 1--10 keV band is shown in Fig. \ref{light}
for the combined SIS instruments. An almost identical light curve is obtained 
for the combined GIS instruments demonstrating that, at least in terms of
the broad band variability, the two SIS and GIS instruments produce
very similar results. During the $\sim 13$ day observation
the source shows a significant brightening by a factor of more than 3
followed by a comparable decline. The fastest variation is a flux increase
of $\sim$ 35\% in roughly 12 hrs. Fig. \ref{light} also shows a plot of the
2--5 keV/5--10 keV softness ratio versus time (for simplicity we show only
the SIS data).  The brightening of the source is clearly associated with a 
significant spectral softening, although the spectrum remains relatively soft 
during the subsequent decline in flux. 

\begin{figure}
\centering
\begin{minipage}{85 mm} 
\centering\hbox{\includegraphics[width=6 cm, angle=270]{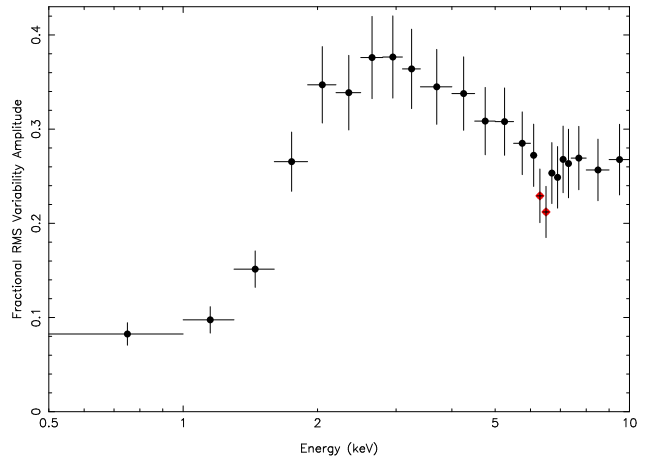}}
\caption{The fractional rms variability amplitude as a function of
energy calculated for the entire \asca ~long-look observation (SIS data).
}
\label{RMS}
\end{minipage}
\end{figure}

We have also investigated the spectral variability exhibited by NGC 4151
by calculating the fractional rms variability amplitude 
as a function of energy (see Edelson \etal 2002). The result, shown in
in Fig. \ref{RMS}, demonstrates, in a model independent fashion, the steep 
rise in the amplitude of the variability above 1 keV to a maximum at $\sim
3$ keV, followed a significant decline at higher energy. There is also a
hint of a dip at the energy of the iron K$\alpha$ line (6.2-6.6 keV).

\section{Fitting the spectral template to the \asca ~data}

In order to study the precise nature of the spectral variability exhibited 
by the source, we have broken the long-look observation up 
into eight sections utilizing gaps in the light curve to define section 
boundaries as illustrated in the lower panel of Fig \ref{light}.
Appropriate response matrices and ancillary response files were constructed 
using standard FTOOLS routines and the spectra were again binned to a 
minimum of 20 counts per bin to allow the use of the
$\chi^{2}$ statistic in the spectral fitting.

Here our goal is to explain the softening of the spectrum as the 
continuum rises and the subsequent lack of spectral evolution
during the continuum decline in terms of photoionization effects. In the 
context of the spectral template description discussed earlier, this implies 
changes in the ionization of the warm absorber induced by variations
in the underlying continuum flux from the central source.

We adopt a very stringent requirement in spectrally fitting the 
\asca ~data in that we fix {\it all but three} of the spectral
parameters of the template model at the values derived from the \bep ~
observation (Table \ref{bsax4-1}).
The three  parameters permitted to vary are the normalization of the
hard power-law continuum, $A_1$, the normalization of the
soft X-ray power-law, $A_2$, and the ionization parameter, $\xi$. 
The column densities of the warm and cold components are kept fixed,
as is the slope of the underlying continuum and the parameters
of the iron K$\alpha$ line. For the Compton reflection, we fixed the
actual component flux at the \bep ~level (rather than fixing the scale 
parameter $R$).

\begin{figure}
\centering
\begin{minipage}{85 mm} 
\centering \hbox{ \includegraphics[width=5.5 cm, angle=270]{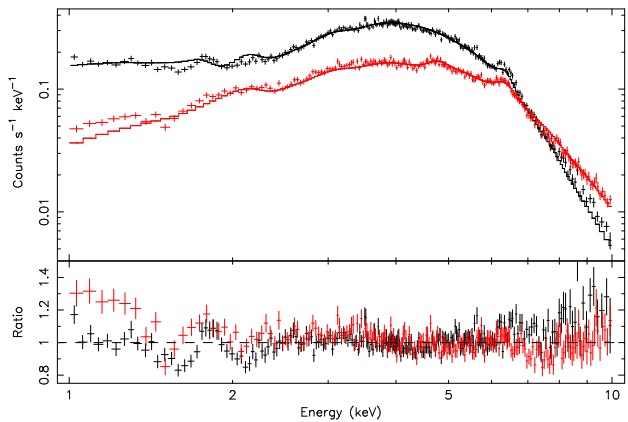}}
\caption{A comparison of the best-fitting spectral template model
to the combined SIS (black) and GIS-2 (red) spectra from time segment 3
of the long-look observation.
{\it Upper panel:} The count rate spectra and best-fit model.
{\it Lower panel:} The ratio of the data to the best-fit model.
Note the discrepancies between the SIS and GIS are most evident 
below $\sim 1.5$ keV and above $8$ keV.}
\label{compare}
\end{minipage}
\end{figure}

Before proceeding with the \asca ~spectral analysis we need to
investigated the degree of inconsistency between the SIS and GIS
instruments. A comparison of the residuals when the above ``template''
model is fitted to the SIS-0/SIS-1 and GIS-2 spectral data from time 
segment 3 is shown in Fig. \ref{compare}. As noted earlier these correspond to
data extracted with a $1'$ radius source cell in the case of
the SIS and a $3'$ radius cell for the GIS. Discrepancies between the
two datasets are most evident below 2 keV and above 8 keV.
The contamination of the GIS spectrum by the BL Lac
appears to account for most of the problem at low
energies. In fact we find that GIS data extracted within a
$1'$ radius cell (thereby greatly reducing the BL Lac contribution)
actually agree reasonably well with the SIS data, although this leads to a 
$\sim 60\%$ loss of counts in the GIS. The problem at the high energy end 
is similar in character to that reported by Turner \etal (2001)
in a comparable \asca ~long-look observation of the narrow-line Seyfert 1 
galaxy Akn 564. Here we
take a somewhat different approach to that employed by
Turner \etal (2001). In the subsequent analysis
we make use {\it solely} of the spectra derived from the SIS data. 
For our investigation the exclusion of the soft BL Lac flux is a
useful advantage which outweighs the spectral uncertainty above
8 keV, since we employ a very tightly constrained model of the hard
continuum (via the adopted spectral template).  In any case,
our main interest is to characterize the spectral changes rather
than to determine precise values for spectral parameters.
Consistency checks show that very similar results to those
reported below for the SIS are obtained with the GIS instruments 
(in fact somewhat better reduced $\chi^{2}$ are recorded) provided
an extra soft component representative of the BL Lac contribution is including 
in the spectral modelling.

\begin{figure}
\centering
\begin{minipage}{85 mm} 
\centering \hbox{ \includegraphics[width=6 cm, angle=270]{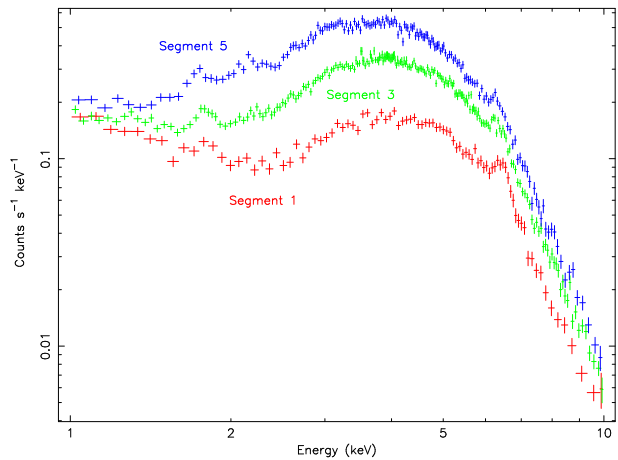}}
\caption{The 1--10 keV combined SIS spectrum of NGC 4151 as measured in three 
time intervals during the \asca ~long-look observation.}
\label{spectra}
\end{minipage}
\end{figure}

Fig. \ref{spectra} compares the SIS spectra from time segments 1, 3 and 5 of 
the long-look observation during which the source count rate in the 1--10
keV band increased by over a factor of 3. These SIS spectra illustrate
that the level of the hard continuum rises by a factor of $\sim 2$
and that it is the marked softening of the spectrum in the 2--6 keV band
that accounts for the additional factor in the count rate increase
({\it cf.} Fig. \ref{RMS}).

\begin{figure}
\centering
\begin{minipage}{85 mm} 
\centering \hbox{ \includegraphics[width=6 cm, angle=270]{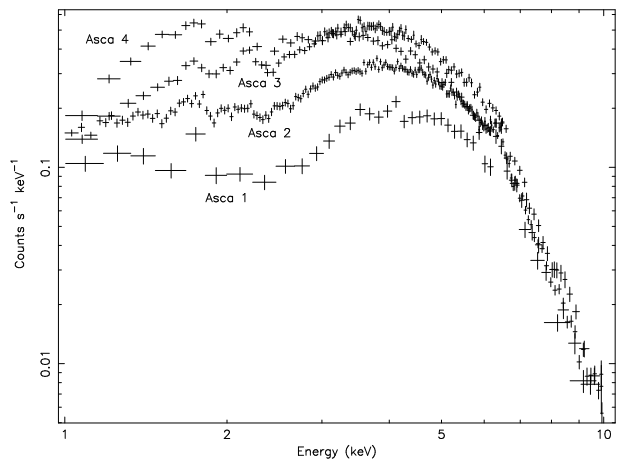}}
\caption{The 1--10 keV X-ray spectrum of NGC 4151 as measured by \asca ~in 
the period 1993-1995. These are count rate spectra from the SIS-0 detector
only.}
\label{asca_past}
\end{minipage}
\end{figure}

The results of fitting the constrained template model to the spectra
from the eight segments of the long-look observation are 
given in Table~\ref{ascallwad}. Overall the model provides a reasonable 
fit to the eight spectral datasets with a combined
$\chi$$^{2}$ of 4616 for 3665 d-o-f (degrees of freedom). 
Interestingly the spectrally variation revealed by the long-look data largely 
encompasses the range of spectral form exhibited by NGC 4151
in earlier \asca ~observations (Fig. \ref{asca_past}). In fact our 
spectral template model provides quite a good description of all the available
datasets from  the previous seven years of observation by both \asca ~and 
\bep, with a range of ionization parameter not dissimilar to that derived 
here.

\begin{table}
\centering
\begin{minipage}{85 mm} 
\caption{Spectral fitting of the eight time segments from the \asca ~
long-look observation.}
\centering
\begin{tabular}{clll}
Segment & log($\xi$) & $A_{1}~^{a}$ & $A_{2}~^{b}$ \\ \hline
1 & 2.483$^{+0.009}_{-0.009}$& 1.99$^{+0.04}_{-0.04}$&
0.82$^{+0.03}_{-0.03}$\\ 
2 & 2.513$^{+0.007}_{-0.005}$& 2.58$^{+0.04}_{-0.04}$&
0.80$^{+0.02}_{-0.02}$\\ 
3 & 2.554$^{+0.006}_{-0.002}$& 3.79$^{+0.03}_{-0.05}$&
0.85$^{+0.02}_{-0.03}$\\
4 & 2.615$^{+0.004}_{-0.013}$& 4.59$^{+0.10}_{-0.01}$&
0.85$^{+0.07}_{-0.02}$\\ 
5 & 2.644$^{+0.001}_{-0.011}$& 5.25$^{+0.08}_{-0.01}$&
0.87$^{+0.06}_{-0.01}$\\ 
6 & 2.636$^{+0.003}_{-0.009}$& 3.50$^{+0.05}_{-0.03}$&
0.88$^{+0.03}_{-0.01}$\\ 
7 & 2.630$^{+0.004}_{-0.009}$& 2.42$^{+0.03}_{-0.02}$&
0.93$^{+0.03}_{-0.03}$\\ 
8 & 2.643$^{+0.006}_{-0.008}$& 2.51$^{+0.04}_{-0.03}$&
0.94$^{+0.04}_{-0.04}$\\ 
\hline
\end{tabular}
\label{ascallwad}
\footnotetext[1]{$10^{-2}\rm~photon~keV^{-1}~cm^{-2}~s^{-1}$ at 1
keV.}
\footnotetext[2]{$10^{-3}\rm~photon~keV^{-1}~cm^{-2}~s^{-1}$ at 1
keV.}
\end{minipage}
\end{table}

If we relax some of the parameter constraints then further improvements
in the fit to the \asca ~long-look segments can be obtained. For example, if 
we allow $E_{K\alpha}$ and 
$I_{K\alpha}$ to vary (but with the parameter value tied across 
the eight spectral datasets), then  we obtain $\chi^{2} = 4552$ for 3663 
d-o-f, 
with best fit values of $E_{K\alpha} = 6.395\pm0.012 $ keV and $I_{K\alpha} 
= 2.2^{+0.2}_{-0.3} \times 10^{-4} \rm photon~cm^{-2}~s^{-1}$ (which according to the F-test is significant at greater than 5 sigma). As a further 
step if we allow the two column density components to vary
(with respect to the \bep ~values) then $\chi^{2} = 4295$ for 3661 d-o-f
with N$_{H,warm} = 2.0\pm0.1 \times 10^{23}\rm~cm^{-2}$ and N$_{H,cold} 
= 4.6 \pm 0.2 \times 10^{22}\rm~cm^{-2}$ (which is very highly significant according to the F-test). In all these fits the largest 
residuals are in range 1--2 keV, where the simple power-law model of the 
soft emission components is clearly a gross over-simplification
(Ogle \etal 2000). However, the presence of these 1-2 keV residuals is very unlikely to influence our interpretation of the observed spectral variations in terms of a variable warm absorber (see below).

\begin{figure}
\centering
\begin{minipage}{85 mm}
\centering 
\includegraphics[width=6 cm, angle=270]{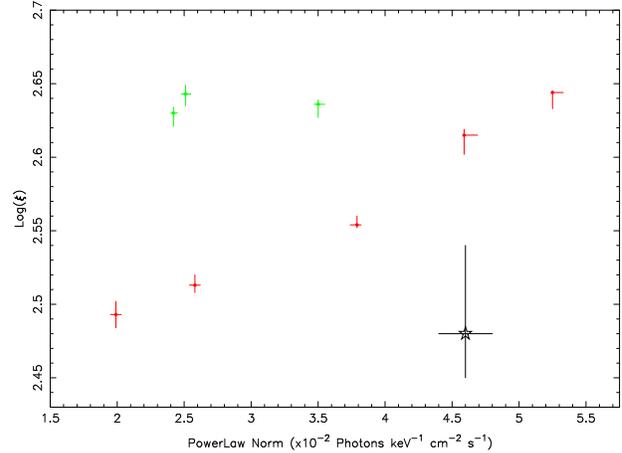}
\caption{Correlation between ionization parameter ($\xi$) and
the normalization of the hard power-law continuum for the eight
segments of the \asca ~``long-look'' observation. The red points mark the segments where the lightcurve is rising, the green
points mark the segments where the lightcurve is falling.
The point marked by the star symbol is from the \bep ~observation;
the error bar is larger in this case due to the additional free parameters
in the spectral fitting. 
}
\label{ioncorr}
\end{minipage}
\end{figure}

The spectral fitting of the \asca ~long-look observation reveals a range
of ionization states for the warm absorber in NGC 4151 (Table~\ref{ascallwad}),
A clear correlation is apparent between the ionization parameter ($\xi$) and 
the level of the hard continuum ($A_{1}$) for the first five segments of the 
observation, which correspond to the period of increasing flux in the light 
curve (Fig~\ref{ioncorr}). However as the flux drops (segments 6--8) the 
correlation is lost and the ionization state remains at a comparable level to 
that seen at the peak of the lightcurve. In contrast, in the \bep ~
observation, although the continuum was at a level comparable to 
that at the peak of \asca ~light curve, in this case the ionization of
the warm absorber was at the bottom end of the range observed with \asca ~
(Fig~\ref{ioncorr}). The observed 
correlation in the \asca ~long-look data lends support to the warm absorber 
interpretation for the observed absorbing column and we argue in \S 5 that 
the lack of a sustained correlation between ionization level and the 
continuum flux, is not necessarily inconsistent with this description. 

The normalization of the soft power-law component, $A_2$, varies by 
$\sim15\%$ during the \asca ~long-look observation. There is also an
apparent factor 2 decline in the soft component over the 18 month interval 
between the \bep ~and \asca ~observations, although as noted earlier
there may be some preferential contamination of the \bep ~data by the nearby 
BL Lac object. Bearing in mind that much of the soft emission in NGC 4151 
originates in the spatially extended component recently imaged by \cha, 
(Ogle \etal 2000; Yang \etal 2001), these results would suggest that there 
may be some leakage of the hard continuum flux through the absorber down to 
$\sim 1$ keV (as suggested by Fig~\ref{RMS}) 
whereas in the current spectral modelling the cut-off is nearer to 
2 keV (see Fig~\ref{model}). Since the complex absorption in 
NGC 4151 is likely to arise from gas exhibiting a range of ionization 
parameters, our two component (cold + warm gas) model must, at best, represent
only a rough approximation to the real system.

\section{Discussion}

\subsection{The physical state of the photoionized absorber in NGC 4151}

One of the major advantages of the warm absorber description is that it is 
based on a physical model which is, in principle, readily testable. 
For example,  given sufficient spectral resolution, sensitivity and dynamic 
range, the true ionization state of the warm gas can be determined 
directly from the absorption edges present in the X-ray spectrum, A further
key diagnostic, somewhat more in tune with current measurement capabilities
in the case of heavily absorbed sources such as NGC 4151, 
is that ionization parameter of the warm gas should track the intensity of 
the ionizing radiation field, albeit in a complex fashion (Nicastro 
\etal 1999). In contrast the popular partial covering (or dual absorber) 
models are based on rather arbitrary, constructed geometries which are very 
difficult to verify.

Previous studies of NGC 4151  have failed to demonstrate in an unambiguous 
fashion that the absorbing medium in NGC 4151 does respond to changes in the 
level of the incident hard continuum (Yaqoob \etal 1989; Fiore \etal 
1990; Warwick \etal 1996; Piro \etal 2002). However, the \asca ~
long-look observation provides an excellent opportunity to study the 
response of the absorber to continuum changes. The correlation 
between ionization parameter and the level of the hard continuum apparent over
the rising part of the light curve (Fig.~\ref{ioncorr}) would appear to be
the signature of a partially photoionized medium. However, a  problem then 
presents itself in that the relatively high ionization state is maintained 
over the latter part of the observation despite the decline in the continuum. 
A possible explanation for this behaviour is that the timescale on which
significant continuum variations are occurring  ($ \sim 3 \times 10^{5}$ s) 
is somewhat shorter than the timescale on which the balance between 
photoionization and recombination can be re-established in the plasma. The
result is that with quasi-continuous continuum variations (of which the
light curve in Fig. \ref{light} represents only a brief snapshot)
the warm absorber is invariably in a non-equilibrium ionization state. 
Nicastro  \etal (1999) have discussed some of the complications which arise
in such circumstances and show that the delayed response of a
medium to sharp increases and decreases in the incident ionizing flux
can give rise to variations in the ionization state apparently unconnected
with flux changes (and even anti-correlations of ionization 
level with flux).
In the case of NGC 4151 there is clear evidence that the medium is slow to 
respond; for example even during the period when the ionization parameter 
and the continuum show a correlation,  the former changes by a factor 
1.8 compared to a factor of 2.5 for the latter. However,
the fact that the degree of ionization tracks the light curve better during 
the rising  segment than during the decline is consistent with the fact that 
$T_{ion} \propto L_{ion}^{-1}$, as is evident from Fig. 1 of Nicastro \etal
(1999). 

The parameter values for the warm absorber in NGC 4151 are not too
far removed from one of the standard cases considered by Krolik \& Kriss
(2001) (although the assumed form of underlying continuum is somewhat 
modified with respect to the version in Krolik \& Kriss \etal (1995), 
which we have adopted in the present work). For example, 
taking $\xi = 300$ and $L_{ion} = 
10^{44} \rm~erg~s^{-1}$ as representative values for the warm absorber in
NGC 4151 and taking the distance of the warm medium from the continuum 
source to be 1 pc, then the inferred gas density is $n \approx 4 \times 10^{4} 
\rm~cm^{-3}$. Table 1 in Krolik \& Kriss (2001) gives the corresponding
equilibration timescales for selected ions in such a plasma. The warm absorber
cut-off is best measured between 2--6 keV, where absorption
edges due to the hydrogen- and helium-like ions of Mg, Si, S, Ar and Ca 
will be of importance (although not individually resolvable in the \asca ~
data). For these ions the relevant equilibration timescales are typically 
between $1-8 \times 10^{6}$ s.  In fact, the measured column density 
for the warm absorber in NGC 4151 of $\sim 2 \times 10^{23} \rm~cm^{-2}$ 
implies $r_{max} {\sim} 0.5$ pc (Krolik \& Kriss 2001), which in turn
(since $t_{ion} \propto r^{2}$), requires a factor 4 reduction in the 
equilibration timescales quoted by Krolik \& Kriss.  Nevertheless it 
is still true that the variability timescale in NGC 4151 is short enough 
for the warm absorber to be out of ionization equilibrium as required.
Since different ions reach their equilibrium abundance, for a given flux of 
ionizing radiation, on different timescales, the detailed form of the 
absorption cut-off will depended on the luminosity history of the source 
stretching back over a period equal to the longest relevant equilibration 
timescale, which in NGC 4151 will be $\sim 10^{7}$ s.

\subsection{The properties of the iron K$\alpha$ line}

In the present paper we have modelled the iron K$\alpha$ emission in a very 
simple fashion, namely as a single narrow line at an energy appropriate
to cold ({\it i.e.} neutral or low-ionization) material.  In the long-look
\asca ~observation, the measured line flux was $2.2^{+0.2}_{-0.3} \times 
10^{-4} \rm~photon~cm^{-2}s^{-1}$, with no evidence for any variations 
in either
the line intensity or in the line profile during the observation period.  
The \bep ~observation gave a somewhat higher value (at the $3\sigma$ level) suggesting
the possibility of line variations, but on the other hand the \asca ~value
is in excellent agreement with the iron K$\alpha$ line flux measured
by \gin ~almost ten years earlier. The lack of significant variability
on timescales of many years suggests the origin of this narrow component 
is in a medium located well outside of the immediate surroundings
of the nuclear source ({\it i.e.} not in a putative accretion disk), but 
probably does not exclude the region occupied by the warm absorber. However, 
the predicted range of iron ionization states in the warm medium is 
Fe XVI -- FeXXI with the associated K$\alpha$ line emission ranging in energy 
from 6.4 -- 6.6 keV. Since the measured line energy excludes much of this 
range, we conclude that the bulk of the line equivalent width (which ranges 
from 82 -- 195 eV during the eight segments of the observation, depending on 
the continuum level) derives from less strongly photoionized material located
further from the nucleus than the warm medium.  The cold absorber in our model
can only account for the observed iron K$\alpha$ flux if it subtends a full
$4\pi$ steradians as viewed from the central source and has an iron
abundance a factor 3 higher than solar.

In contrast with previous claims suggesting a complex broad iron line 
structure in NGC 4151 (Yaqoob, \etal 1995; Wang \etal 1999, 2001; Yaqoob, 
\etal 2001), we find no compelling evidence to include a relativistically 
broadened line in any of our modelling (the best fit intrinsic line width 
of the ``narrow'' line in the \asca ~data was $110 \pm{20}$ eV but this may 
over estimate the true line width given the SIS calibration uncertainties).
The lack of any variability in the line profile during the observation 
contrasts with earlier claims of line profile variability in NGC 4151 
on $10^{4}$ s timescales (Wang \etal 2001). 

Clearly a very extended broad-line profile is easily confused with a
continuum form including a non-negligible contribution from Compton 
reflection and modified by a complex absorption. The complex variable 
nature of the underlying continuum and the absorption may then easily 
manifest itself as variability in the broad-line profile. We will return 
to this issue at a later date utilizing newly acquired data from \xmm. 

\subsection{The spectral index versus flux correlation}

\begin{figure}
\centering
\begin{minipage}{85 mm} 
\centering\hbox{\includegraphics[width=6 cm, angle=270]{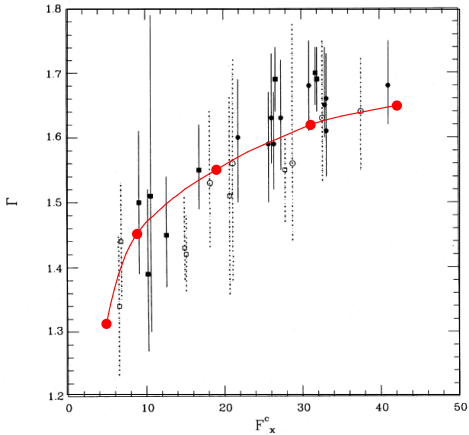}}
\caption{Measurements of the spectral photon index versus the 
absorption-corrected 2-10 keV flux from \exo ~and \gin ~observations,
taken from Fig.~2 of Yaqoob \etal (1993). The solid curve represents the 
results of a simulation in which a constant Compton-reflection 
component was included along with a power-law continuum with 
a variable normalization but fixed slope ({\it i.e.,} $\Gamma = 1.65$).
See the text for further details.}
\label{ginga}
\end{minipage}
\end{figure}

Previous observations of NGC 4151 with both \exo ~and \gin ~have revealed an 
apparent correlation  between spectral index measured in the 2--10 keV band 
and the level of the underlying  continuum 
(Perola \etal 1986; Yaqoob \& Warwick 1991; Yaqoob \etal 1993). Fig. 
\ref{ginga} shows a compilation of measurements from \exo ~and \gin ~(taken 
from Yaqoob \etal 1993) which suggests that as the source brightens the photon
index steepens, from $\Gamma~\sim~1.35$ to $\Gamma~\sim~1.7$, at which
point the correlation saturates.
Yaqoob \etal (1993) also note that the spectral slope changes occur on 
roughly the same timescale as the flux variations. However, a perplexing 
aspect of the spectral index flux correlation observed in the 
medium energy bandpass is that the effect is not mirrored in the 
hard X-ray/soft gamma regime (Johnson \etal 1997; Piro \etal 2002).

An underlying assumption of the spectral template model adopted in the present
paper is that the spectral index of the hard power-law continuum in NGC 4151
remains constant (at a value $\Gamma \approx 1.65$), as the continuum
flux changes, in apparent contradiction to the correlation noted above. (Testing this assumption against the \asca ~observation demonstrates that the \asca ~datasets are broadly consistent with the assumed value of $\Gamma$=1.65, however the values are poorly constrained due to the very limited effective bandpass, and the fact that the values of $\Gamma$ derived are very dependent on the exact details of the applied spectral model).

We have investigated the possibility that the spectral index versus 
relation may actually be a misinterpretation
of the softening of the NGC 4151 spectrum as the source flux increases,
bearing in mind that the reported correlation was established using
both proportional counter data with limited spectral resolution and 
a highly simplified spectral model (in fact just a power-law 
continuum  model, modified by a heavy cold absorbing column).  To this end
we have taken our spectral template model and an appropriate \gin ~
response, and simulated a set of five \gin ~spectra using  XSPEC. In this
simulation, the normalization of the hard power-law continuum was increased 
in steps from $1$ to $10 \times 10^{-2} \rm~photons~cm^{-2}~s^{-1}$,
but with all the other parameter values, (including the ionization 
parameter and the {\it flux} in the Compton-reflection component) frozen at 
the values obtained from the \bep ~spectral fitting.
The simulated 3.5--20 keV \gin ~spectra were then each 
fitted with a simple absorbed power-law model ({\it cf.} Yaqoob \& Warwick 
1991; Yaqoob \etal 1993) and the effective photon spectral index measured. 
The result was a spectral index versus 2--10 keV flux correlation
closely matching the reported correlation (Fig.~\ref{ginga}). Since the
simulation did not include any of the complications associated with varying
ionization parameter of the warm absorber, it follows that the observed
spectral changes are induced solely as a result of the presence of a 
non-varying spectrally-hard Compton-reflection component. Clearly, in this 
scenario, as the direct continuum brightens, the relative contribution of the 
Compton-reflection declines and  the overall spectrum softens until 
eventually the measured spectral index saturates at the value pertaining
to the underlying continuum.

\section{Conclusion}

There has been much debate as to whether NGC 4151 is simply a very
bright, nearby representative of the broad class of galaxies harbouring 
luminous Seyfert nuclei or a unique object (Ulrich 2000). In the X-ray band 
this uncertainty is due, in large measure, to the heavy, complex and variable 
absorption which characterizes the 2--6 keV spectrum of NGC 4151 but 
which is not 
mirrored in the great majority of well-studied Seyfert galaxies 
(Mrk 6 is arguably the best known example of a NGC 4151 analogue, 
Feldmeier \etal 1999).

In this paper we have presented a detailed analysis of two archival 
datasets of NGC 4151, one from \bep ~the other from \asca, which tests the 
hypothesis that the complex absorption is due to a combination of
warm and cold gas distributed along the line of sight to the active 
Seyfert nucleus in this source.  We find that this relatively simple model
does indeed provide a good description of the two datasets in question
and, in fact, gives a reasonable fit to all of the \asca ~and \bep ~
observations from the past seven years. Our key finding is that the warm 
absorber corresponds to a relative low density ($\sim 10^{5}$ cm$^{-3}$) gas, 
with a column density of $\sim 2 \times 10^{23} \rm~cm^{-2}$ and 
with ionization parameter in the range log($\xi$) $\approx 2.4-2.7$.
Since the equilibration timescale for the dominant ions is of the same
order or longer than the timescale of the continuum variability,
the warm component is invariably observed in a non-equilibrium ionization 
state. Non-equilibrium conditions help explain why past 
studies based on relatively short observations spaced by weeks, months or 
years have failed to identify a consistent signature of the warm absorber 
in this source.

In a recent paper,  Krolik \& Kriss (2001) discuss how evaporation from the 
inner edge of the obscuring torus in an AGN can give rise to an 
inhomogeneous photoionized wind in which a broad range of temperatures 
coexist in equilibrium. Krolik \& Kriss (2001) then go on to suggest that this 
wind is the origin of the highly ionized, warm absorbers seen in over half 
of type 1 Seyfert galaxies. For most Seyfert 1 galaxies the properties of
the warm absorber are inferred from the OVII and OVIII absorption edges, 
whereas in NGC 4151 the combination of warm and cold absorbers
gives rise to a sharp spectral cutoff below $\sim 2$ keV, thus eliminating 
the oxygen features.  Nevertheless, as noted earlier, the inferred
properties of the warm component in NGC 4151 are not too far removed
from one the standard cases detailed by Krolik \& Kriss (2001). It is
therefore reasonable to hypothesise that the warm absorber in NGC 4151
originates in such a multi-temperature wind, but that the unusually
complex character of the absorption is due to a line of sight
which grazes the top edge of the obscuring torus so as to intercept
a substantial column of {\it both} the warm and cold components. In essence
in terms of its X-ray absorption properties, NGC 4151 is intermediate 
between the Seyfert galaxies in which the absorption, if present, is 
predominantly due to the Krolik \& Kriss's multi-temperature wind 
(type 1 objects) and those where the cooler material of the obscuring 
torus dominates (type 2 objects).  

The other findings of this paper are that (i) the reported hardening of 
the spectrum of NGC 4151 as the continuum level falls may be simply due to 
the presence of an underlying (hard and relatively constant) 
Compton-reflection component and (ii) there is no compelling 
evidence for a relativistically broadened iron K$\alpha$ line in
NGC 4151 contrary to earlier claims. 

In summary, it appears that many of the X-ray properties of NGC 4151 can 
be explained in terms of the current paradigm for Seyfert galaxies.
Future observations with \xmm, \cha ~and \int ~will be able to test
the degree to which our spectral template description is valid
and should further illuminate the question of whether NGC 4151 should
be regarded as the archetypal Seyfert galaxy.

\vspace{1cm}

{\noindent \bf ACKNOWLEDGMENTS}

\vspace{2mm}

NJS gratefully acknowledges the financial support from PPARC. It is a 
pleasure to thank Simon Vaughan for his many valuable discussions, and 
Keith Arnaud for his help with XSPEC The archival \asca ~X-ray data used 
in this work was obtained from the
Leicester Database and Archive Service (LEDAS) at the Department of
Physics \& Astronomy, University of Leicester. The archival \bep ~
X-ray data used in this work was obtained from the \bep ~ASI Science 
Data Center at ESRIN, Italy. This research has made extensive use of 
NASA's Astrophysics Data System Abstract Service.


\end{document}